# Rare-Earth Control of the Superconducting Upper Critical Field in Infinite-Layer Nickelates

**Authors:** Bai Yang Wang[1,2]*†, Tiffany C. Wang[1,3]*†, Yu-Te Hsu[4], Motoki Osada[1,5], Kyuho Lee[1,2], Chunjing Jia[1], Caitlin Duffy[4], Danfeng Li[1], Jennifer Fowlie[1,3], Malcolm R. Beasley[3], Thomas P. Devereaux[1,5], Ian R. Fisher[1,3], Nigel E. Hussey[4,6], Harold Y. Hwang[1,3]*

**Affiliations:**

[1]Stanford Institute for Materials and Energy Sciences, SLAC National Accelerator Laboratory; Menlo Park, CA 94025, United States.

[2]Department of Physics, Stanford University; Stanford, CA 94305, United States.

[3]Department of Applied Physics, Stanford University; Stanford, CA 94305, United States.

[4]High Field Magnet Laboratory (HFML-EMFL) and Institute for Molecules and Materials, Radboud University; Toernooiveld 7, 6525 ED Nijmegen, Netherlands.

[5]Department of Materials Science and Engineering, Stanford University; Stanford, CA 94305, United States.

[6]H. H. Wills Physics Laboratory, University of Bristol; Tyndall Avenue, Bristol BS8 1TL, United Kingdom.

*Corresponding author. Email: bwang87@stanford.edu; catwang@stanford.edu; hyhwang@stanford.edu.

†Both authors contributed equally.

**Abstract:**

The consequences of varying the rare-earth element in the superconducting infinite-layer nickelates have been much debated. Here we show striking differences in the magnitude and anisotropy of the superconducting upper critical field across the La-, Pr-, and Nd-nickelates. These distinctions originate from the 4$f$ electron characteristics of the rare-earth ions in the lattice: they are absent for La$^{3+}$, nonmagnetic for the Pr$^{3+}$ singlet ground state, and magnetic for the Nd$^{3+}$ Kramer's doublet. The unique polar and azimuthal angle-dependent magnetoresistance found in the Nd-nickelates can be understood to arise from the magnetic contribution of the Nd$^{3+}$ 4$f$ moments. In the absence of rare-earth effects, we find that the nickelates broadly violate the Pauli limit. Such robust and tunable superconductivity suggests potential in future high-field applications.

**Main Text:**

Following the initial discovery of superconductivity in $Nd_{0.8}Sr_{0.2}NiO_2$ (*1*), the subsequent development of rare-earth (*R*) variants with Pr and La in place of Nd (*2–4*) has created an emerging new family of superconducting infinite-layer nickelates (Fig. 1A). Unlike the cuprates, with which the nickelates are often compared, *R*-site variations are expected to have impact beyond ionic size effects. There is consistent transport, spectroscopic, and theoretical evidence for hybridization between the *R* 5*d* and the Ni 3*d* orbitals (*2–15*). An open question currently under debate is the potential role of the *R* 4*f* electrons (*16–19*). Here we report that the *R*-site directly controls the magnitude and anisotropy of the superconducting upper critical field $H_{c2}$.

Figure 1 shows a central result motivating this study – that the magnetic field suppression of superconductivity differs widely among *R*-variants, which does not scale with $T_{c0}$, the superconducting transition temperature in zero external magnetic field. Magnetoresistance measurements up to 35 T of representative (*R*,Sr)NiO$_2$ (*R* = La, Pr, Nd) samples at optimal doping are shown in Fig. 1, B-D, for both out-of-plane ($H_\perp$) and in-plane ($H_\parallel$) field orientations. The corresponding resistively determined upper critical fields ($H_{c2\perp}$ and $H_{c2\parallel}$; defined as 50% of the normal state resistivity) are shown (Fig. 1E), where the drastically different scales for $H_{c2}$ are evident.

For a superconductor in the dirty limit, which describes the nickelates (*20*), the effective Ginzburg-Landau coherence length $\xi$ is the geometric mean of the Pippard coherence length $\xi_0$ and the mean-free-path (*21*). Then, as $H_{c2\perp} = \frac{\phi_0}{2\pi\xi^2}(1 - \frac{T}{T_{c0}})$, $H_{c2\perp} \propto 1/\xi_0 \propto \Delta_0 \propto T_{c0}$. Here $\phi_0$ is the magnetic flux quantum and $\Delta_0$ the superconducting gap. As for $H_{c2\parallel}$, the observed $(1-T/T_{c0})^{1/2}$ temperature dependence indicates dominant paramagnetic de-pairing near $T_{c0}$ (20). $H_{c2\parallel}$ is also proportional to $T_{c0}$ at temperatures near $T_{c0}$, described by $-\ln\left(\frac{T}{T_{c0}}\right) = 7\eta(3)\left(\frac{\mu_0 H_{c2\parallel}}{2\pi k_B T_{c0}}\right)^2$ (22), where $k_B$ is the Boltzmann constant, $\mu_0$ is the vacuum magnetic permeability, and $\eta(x)$ is the Riemann zeta function. Therefore, we normalize both $H_{c2}$ and $T$ by $T_{c0}$ (Fig. 1F) for all samples. We observe a systematic variation in $H_{c2}/T_{c0}$, which is largest for La-nickelates, somewhat smaller for Pr, and significantly reduced for Nd. Furthermore, this variation between $R$-sites is persistent across the respective superconducting domes via Sr doping (Fig. 1F; see figs. S1 and S2 for complete data sets).

To probe the origin of this striking variation in $H_{c2}$, we measure in detail the magnetoresistance as a function of the polar ($\theta$) and azimuthal ($\phi$) angle of the applied magnetic field direction, as illustrated in Fig. 2A. In the normal state, negligible angular dependence is observed for all samples (Fig. 2C and D). Therefore, we focus on the superconducting transition region with finite sample resistance below the normal state value. We first present the characteristic features through representative measurements (Fig. 2E; see figs. S3-S5 for complete data sets). The La- and Pr-nickelates both show conventional angular dependence; their '∞'-shaped $\theta$-dependence shows that superconductivity is more robust to in-plane magnetic fields than perpendicular fields, as commonly seen in layered superconductors (23). Their 'figure-of-eight'-shaped $\phi$-dependence, arising from vortex motion, also follows previous studies of other type-II superconductors (23, 24). In sharp contrast, the Nd-nickelates show qualitatively distinct magnetoresistance. The orientation of the $\theta$-dependence is opposite to that of La-/Pr-nickelates and other layered superconductors, - i.e. superconductivity is *more* suppressed by an in-plane magnetic field, indicating a complete reversal of the superconducting anisotropy. Furthermore, a striking clover-leaf $\phi$-dependence is observed for Nd. As shown in fig. S6, we can rule out a vortex-related origin of these features.

How can we understand this clear difference between the Nd-nickelates versus the La-/Pr-nickelates? In all cases the in-plane lattice parameters are clamped to that of the SrTiO$_3$ substrate, while the $c$-axis variations reflect the $R$-site ionic radii (*1–3*): La$^{3+}$ is the largest, ~4.5% larger than the closely similar Pr$^{3+}$ and Nd$^{3+}$ (within 1.5% of each other) (*25*). Thus cation size effects do not seem to account for the unique $H_{c2}$ features in Nd. Rather, the clover-leaf $\phi$-dependent magnetoresistance in Nd is more suggestive of a potential role of local magnetic moments, leading to a consideration of the 4$f$ electrons and their crystal electric field (CEF) structure.

To examine this, we use theoretical approaches including *ab initio* density functional theory (DFT) calculations (*26, 27*), Bader charge analysis (*28*), and symmetry arguments to determine the 4$f$ levels split by the tetragonal CEF (see supplementary materials for details). In Fig. 3A, we show the electron density iso-surface of the Ni 3$d$, 4$s$, O 2$p$, and $R$ 5$d$, 6$s$ orbital electrons within a unit cell where the contours correspond to 2.5% of the maximum electron density. Despite the appreciable hybridization, especially between the O 2$p$ and Ni 3$d$ orbitals, most of the electron density remains within the vicinity of the parent ion. This motivates the use of a point charge approximation in determining the CEF-split 4$f$ energy levels, as shown for Pr$^{3+}$ and Nd$^{3+}$ as a function of $c$-axis lattice constant (Fig. 3, B and C), where the range of experimental values is highlighted by the vertical blue bands. While La$^{3+}$ has no 4$f$ electrons, we find a singlet crystal field ground state for Pr$^{3+}$ which to first order is also nonmagnetic. By contrast, Nd$^{3+}$ has a doublet ground state as expected for a Kramer's ion, making this the lone magnetic $R$-site ion among the three. Furthermore, by adding an external field term to the CEF perturbation Hamiltonian, we find that the Nd 4*f* moment shows an easy-plane magnetic anisotropy (fig. S7).

Following these CEF considerations, we can construct a simple model that quantitatively describes all of the angular magnetoresistance data, based on the assumption that the Nd 4*f* moment contributes purely to the background magnetization and magnetic permeability of the lattice. For the $\theta$-dependent magnetoresistance, all of the nickelates have in common anisotropic orbital depairing (AOD; Fig. 4A) as a natural consequence of the shorter *c*-axis coherence length (*21*), and hence an $H_{c2\|}$ larger than $H_{c2\perp}$. This can be seen to dominate at all temperatures for La-/Pr-nickelates (fig. S3). For Nd-nickelates, there is an additional competing term due to the enhanced magnetic permeability (EMP; Fig. 4B) arising from the 4*f* moments, which has crucially the *opposite* anisotropy given the easy-plane magnetic anisotropy.

The temperature-dependent balance of this competition (Fig. 4F) underlies the remarkable reversal of the $\theta$-dependent magnetoresistance anisotropy we observe for highly-doped Nd-nickelates (Fig. 4G) – i.e. the temperature-dependent crossing of $H_{c2\perp}$ and $H_{c2\|}$. Our minimal model is surprisingly quantitative; the measured magnetoresistance of all samples (La, Pr, Nd), doping levels, temperatures, and magnetic fields can be described by one generic form for the resistance $R_{AOD}(\theta)$ arising from AOD, and another generic form for the resistance $R_{EMP}(\theta)$ from EMP for Nd, such that the total $\theta$-dependent resistance $R^{PL}(\theta)$ can be expressed as:

$$R^{PL}(\theta) = C + \alpha_{AOD} R_{AOD}(\theta) + \alpha_{EMP} R_{EMP}(\theta). \tag{1}$$

Here $C$ is a constant, and $\alpha_{AOD}$ and $\alpha_{EMP}$ are the temperature- and magnetic-field-dependent coefficients of the two functional forms (fig. S8). The dashed lines in Figs. 4G and S4 show the accuracy of this two-component description. In particular, as the polar anisotropy reverses with temperature, the competing AOD and EMP terms produce complex $\theta$-dependences with up to 6 local resistance maxima (fig. S4) which can be well described by Eq. 1.

As for the $\phi$-dependent magnetoresistance, we see a competition between a 2-fold symmetric ($C_2$) $\phi$-dependence and a 4-fold symmetric ($C_4$) $\phi$-dependence (Fig. 4D, E, and H). The $C_2$ response has been seen previously (23, 24) and is a generic result of the $\phi$-dependence of the Lorentz force exerted by the measurement current on vortices, since only the perpendicular component contributes. Thus this $\phi$-dependence is seen in all $R$-variants of the nickelates. However, uniquely in the Nd-nickelates, we observe an additional $\sin(4\phi)$ component (clover-leaf pattern, Fig. 4I), such that the total $\phi$-dependent resistance $R^{AZ}(\phi)$ can be then expressed as:

$$R^{AZ}(\phi) = C + \alpha_{c2} \sin(2(\phi - \phi_{c2})) + \alpha_{c4} \sin(4(\phi - \phi_{c4})). \qquad (2)$$

Here $C$, $\phi_{c2}$, and $\phi_{c4}$ are constants, and $\alpha_{c2}$ and $\alpha_{c4}$ are the temperature- and magnetic-field-dependent coefficients (fig. S5 and S8). We attribute the $C_4$ symmetric dependence again to the $4f$ moments, since the occupied CEF levels are necessarily invariant to $C_4$ symmetry operations. This implies an in-plane magnetic easy axis along the [110] and equivalent directions. This interpretation is bolstered by the experimental observation that $\phi_{c2}$ follows the applied current direction, while $\phi_{c4}$ is locked to the lattice (fig. S9).

The extracted coefficients $\alpha_{AOD}$, $\alpha_{EMP}$, $\alpha_{c2}$, and $\alpha_{c4}$ exhibit peaks in magnetic field for a given constant temperature (fig. S8), which is a natural result of the boundary conditions: in either the fully superconducting or normal state, no angle-dependent magnetoresistance is seen. With decreasing temperature, both the peaks in $\alpha_{EMP}$ and $\alpha_{AOD}$ shift to higher magnetic fields, tracing the superconducting phase boundary in the temperature-field plane (Fig. 2B). However, the magnitude of $\alpha_{AOD}$ drops significantly with decreasing temperature, contrasting with the increase in $\alpha_{EMP}$.

The pair-breaking framework of superconductivity is useful to discuss these observations. To leading order, the external magnetic field suppresses superconductivity through either orbital or paramagnetic de-pairing (21). While orbital de-pairing has an intrinsic $\theta$-dependence rooted in the superconducting anisotropy, paramagnetic de-pairing is largely isotropic as it suppresses superconductivity by the Zeeman energy cost. Therefore, although both effects are enhanced by the local magnetic field arising from the Nd 4$f$ moments, an underlying $\theta$-dependence of the local magnetic field intensity is more readily visible when paramagnetic de-pairing is dominant. With this in mind, we can associate $\alpha_{AOD} R_{AOD}(\theta)$ with orbital de-pairing, and $\alpha_{EMP} R_{EMP}(\theta)$ with paramagnetic de-pairing which dominates at low temperatures.

In contrast, neither de-pairing effect has intrinsic $\phi$-dependence, and both would faithfully exhibit the structure of the local magnetic field. Therefore, the clover leaf pattern is a direct manifestation of the $\phi$-dependence of the local magnetic field intensity, particularly since $\alpha_{c4}$ is consistently an order of magnitude larger than $\alpha_{c2}$ where it is observed (fig. S8). With decreasing temperature, the onset and dramatic increase of $\alpha_{c4}$ signal the growing contribution of the 4$f$ moments, which in turn enhances $\alpha_{EMP}$. Note that while the magnetic response of the 4$f$ moments likely exists over a broader range of temperature and magnetic field, it only appears in the magnetoresistance in the vicinity of $H_{c2}$ through pair-breaking effects, and in this process becomes intertwined with the superconducting response. Therefore, the exact temperature or field dependence of the 4$f$ magnetic moment magnetization cannot be simply extracted from these data. However, the impact of Nd on $H_{c2}$ is suggestive of low temperature Nd 4*f* antiferromagnetic exchange, as is typically observed for complex materials hosting magnetic *R*-sites (*29*) (figs. S10).

With these developments, we finally return to the significant differences in $H_{c2}$ across *R*-variants shown in Fig. 1. With important 4*f* moment enhancements of the field, Nd has the lowest $H_{c2}/T_{c0}$; in their absence, $H_{c2}/T_{c0}$ is highest for La. Pr is intermediate, consistent with the contribution of a $Pr^{3+}$ 4*f* moment induced by magnetic field. Thus we see that the underlying intrinsic behavior of $H_{c2}$ appears to be similar across the infinite-layer nickelates, once the additional suppression due to the *R*-site magnetism has been taken into account. This leads to the conclusion that the conventional weak-coupling Pauli limit $H_{c2} = 1.86\, T_{c0}$ (*30*) is broadly exceeded. Although the (Nd,Sr)NiO$_2$ samples obey this limit (Fig. 1F) (*20*), we now understand this to occur due to Nd 4*f* moment contributions. We observe that the Pauli limit is surpassed by up to a factor of 3.5 in the La series (Fig. 1F). Note that we use a 50% resistance criterion for $H_{c2}$, to minimize contributions from superconducting fluctuations and vortex creep (*20*); adopting higher resistance thresholds would make this Pauli-limit violation even larger.

While the magnitude of $H_{c2}$ presents an important question for future research, we discuss potentially relevant scenarios. We observe that the angular magnetoresistance in the Nd-nickelates can be fully captured by considering only the 4*f* moment contribution to the background magnetization and magnetic permeability. However, it is important to note that this does not preclude other contributions from the hybridization between the *R* 5*d* and the Ni 3*d* orbitals – in the presence of finite spin-orbit coupling, multi-orbital mixing could enhance $H_{c2}$. Perhaps the most straightforward explanation is that the experimentally observed $T_{c0}$ via transport occurs in the presence of significant phase fluctuations/disorder and does not reflect the pairing scale $\Delta_0$.

Indeed, tunneling spectroscopy indicates a large $\Delta_0$ to $T_{c0}$ ratio, although surface effects and variations remain to be clarified (*31*). Further evidence for this scenario may be found in a recent study of $La_{0.8}Sr_{0.2}NiO_2$ with higher $T_{c0}$ but comparable extrapolations of $H_{c2}$ to what we observe (*32*). The fact that experimentally, nickelate superconductivity is in the dirty limit appears to us to disfavor more exotic mean-field enhancements. Independent of origin, the unusually large $H_{c2}/T_{c0}$ ratios in the nickelates, combined with sensitive control through the *R*-site, promise exciting potential in high-field applications.

**NOTE:**

During the preparation of this manuscript, we became aware of two reports on $H_{c2}$ in La-nickelates (*32, 33*).

**Acknowledgements**

We thank S. A. Kivelson, C. Murthy, S. Raghu, and Yue Yu for fruitful discussions, and Yijun Yu and S. P. Harvey for critical reading of the manuscript. We acknowledge the support of the HFML-RU/NWO, a member of the European Magnetic Field Laboratory (EMFL).

**Funding:**

US Department of Energy, Office of Basic Energy Sciences, Division of Materials Sciences and Engineering grant DE-AC02-76SF00515. (BYW, TCW, MO, KL, CJ, DL, TPD, IRF, HYH)

Gordon and Betty Moore Foundation's Emergent Phenomena in Quantum Systems Initiative grant GBMF9072. (JF, HYH)

Swiss National Science Foundation through Postdoc.Mobility P400P2199297 and Division II 200020 179155. (JF)

European Research Council (ERC) under the European Union's Horizon 2020 research and innovation programme (grant agreement no. 835279-Catch-22) (YTH, CD, NEH)

**Author contributions:**

Conceptualization: BYW, HYH

Sample Synthesis: MO, KL, DL, JF

Transport Characterization: BYW, TCW, YTH, CD

Theoretical Calculation: BYW, CJ

Funding acquisition: HYH, TPD, NEH

Supervision: MRB, TPD, NEH, IRF, HYH

Writing: BYW, TCW, HYH

**Competing interests:**

Authors declare that they have no competing interests.

**Data and materials availability:**

All data are available in the main text or the supplementary materials.

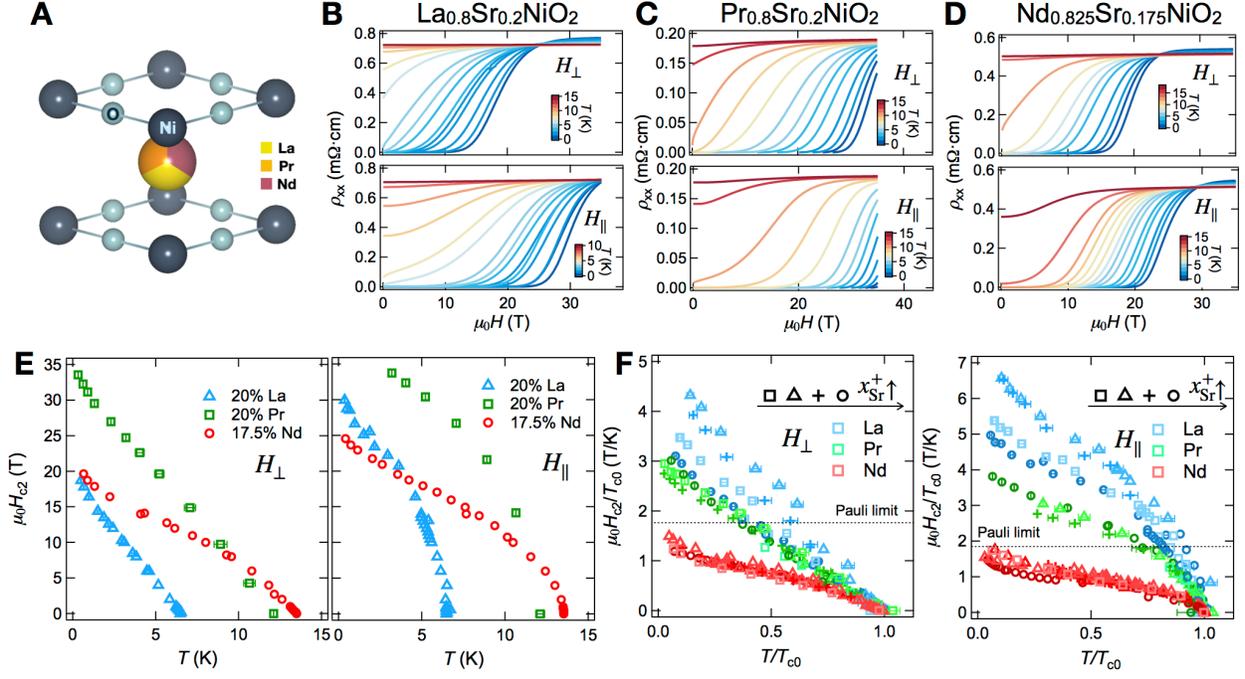

**Fig. 1. Structure, magnetotransport, and $H_{c2}$ properties of $(R,Sr)NiO_2$.** (**A**) Schematic unit cell structure of $(R,Sr)NiO_2$ ($R$ = La, Pr, or Nd; Sr substitution not indicated). (**B**, **C**, **D**) Magnetoresistance of representative samples near optimal doping: $La_{0.8}Sr_{0.2}NiO_2$, $Pr_{0.8}Sr_{0.2}NiO_2$, and $Nd_{0.825}Sr_{0.175}NiO_2$ (*12*) respectively, at temperatures ranging from 0.34 K to 18 K for out-of-plane ($H_\perp$) and in-plane ($H_\parallel$) field orientations. (**E**) Temperature dependence of $H_{c2}$ determined from data shown in B, C, and D. (**F**) Normalized $H_{c2}/T_{c0}$ against reduced temperature $T/T_{c0}$ of $R_{1-x}Sr_xNiO_2$ for $H_\perp$ and $H_\parallel$ orientations, respectively. Here $x$ = 0.15, 0.175, 0.2, 0.225 (*12, 20*) for Nd, $x$ = 0.16, 0.18, 0.20, 0.24 for Pr, and $x$ = 0.15, 0.16, 0.18, 0.20 for La. The dashed lines indicate the Pauli limit of $H_{c2}$ = 1.86 $T_{c0}$.

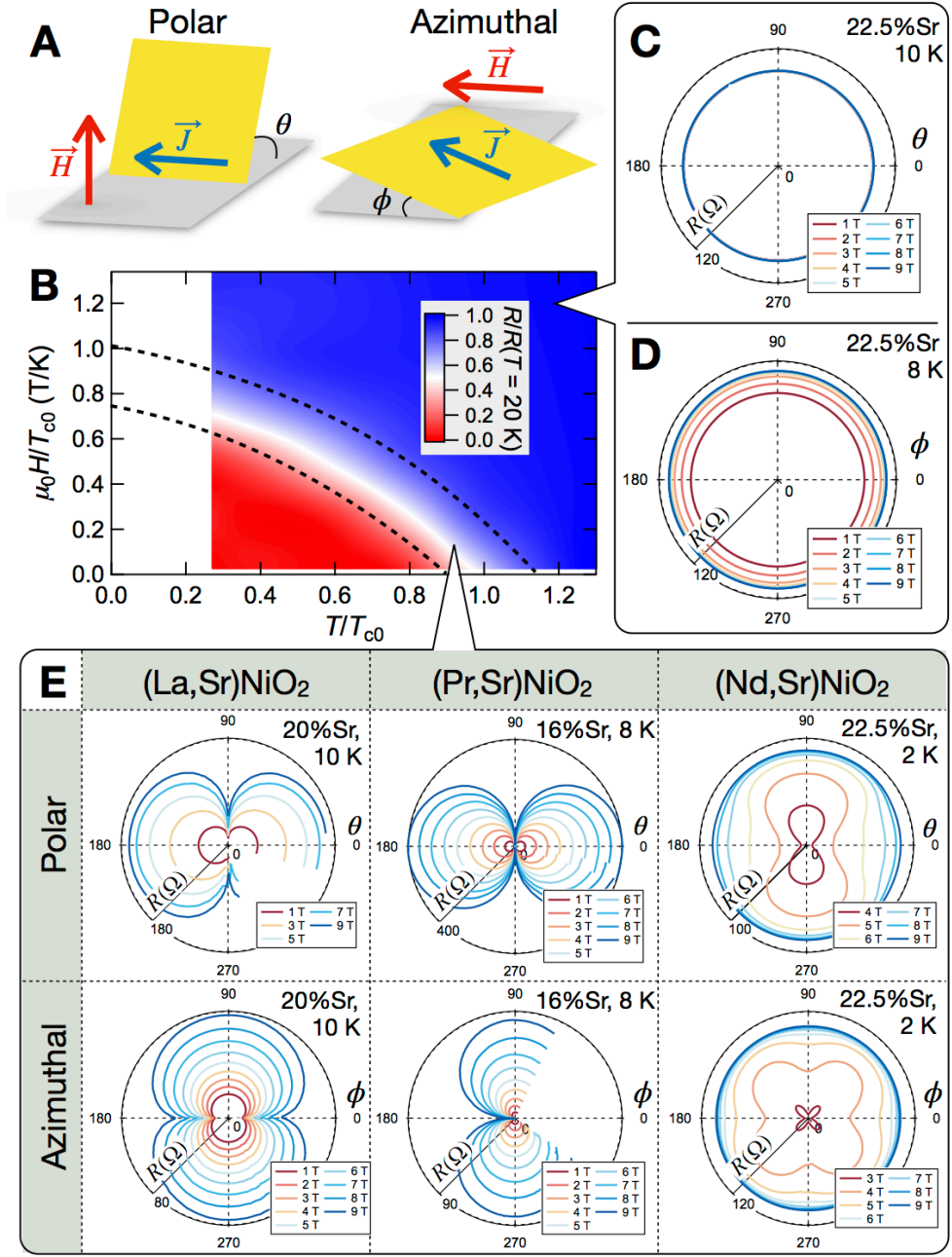

**Fig. 2. Representative characteristics of the angular magnetoresistance in ($R$,Sr)NiO$_2$.** (**A**) Schematic of the measurement geometry with respect to the sample (yellow square). The blue and red arrows represent the directions of the measurement current and the external magnetic field, respectively. (**B**) Magnetic field versus temperature phase diagram of a representative Nd$_{0.775}$Sr$_{0.225}$NiO$_2$ sample ($T_{c0}$ = 7.4 K). (**C** and **D**) Representative polar ($\theta$) and azimuthal ($\phi$)

angle dependence of the magnetoresistance in the normal state for $Nd_{0.775}Sr_{0.225}NiO_2$. (**E**) Representative polar and azimuthal angle dependence of magnetoresistance in the superconducting transition region for La-, Pr-, and Nd-nickelates, with the specific doping and temperature indicated at the top right corner of each panel.

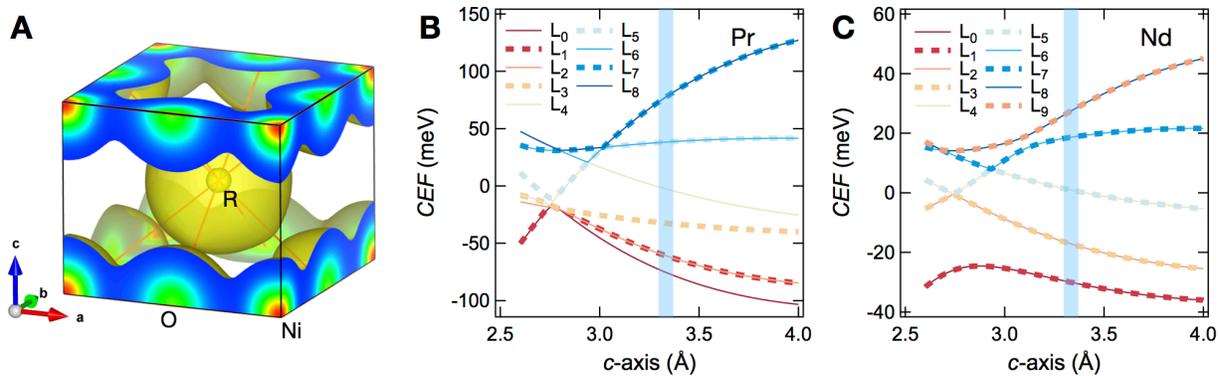

**Fig. 3. Calculation of 4$f$ crystal electric field (CEF) levels.** (**A**) DFT-calculated electron density distribution of NdNiO$_2$, with the valence including Nd 5$d$, 6$s$, Ni 3$d$, 4$s$, and O 2$p$ electrons. The yellow iso-surface gives the contour corresponding to 2.5% of the maximum electron density. DFT calculations using a Pr pseudopotential result in a visually indistinguishable charge distribution. (**B** and **C**) Calculated CEF splitting as a function of $c$-axis lattice constant for Pr- and Nd-nickelates, respectively. The blue shaded region marks the range of experimental $c$-lattice constant values. The crystal field levels L$_x$ are expanded in table S1.

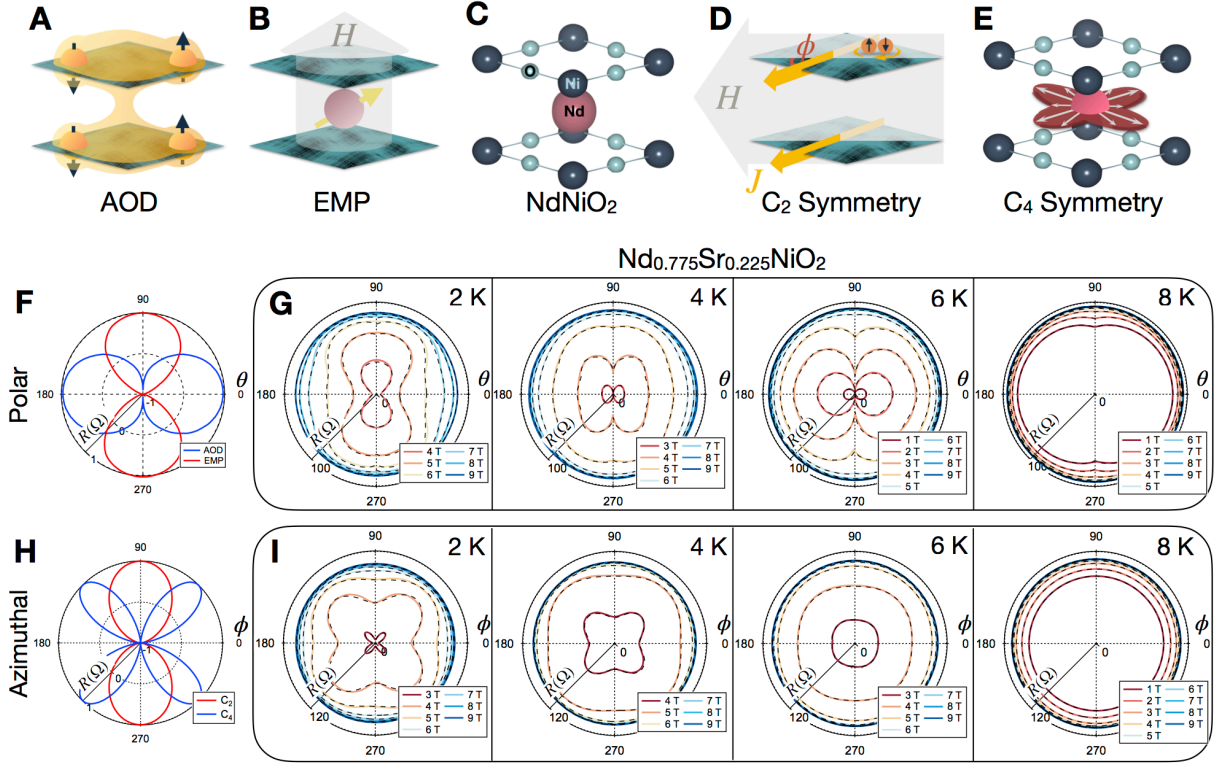

**Fig. 4. Competing effects in the angular magnetoresistance.** (**A** to **E**) Schematic illustrations of (A) anisotropic orbital de-pairing (AOD), (B) enhanced magnetic permeability (EMP), and the origin of the (D) $C_2$ and (E) $C_4$ symmetric azimuthal response. (C) shows the corresponding unit cell. (**F** and **H**) Fitting functions corresponding to the AOD and EMP effects for $\theta$-dependence and $C_2$ and $C_4$ symmetric $\phi$-dependence, respectively. (**G** and **I**) Temperature and magnetic field dependence of the $\theta$- and $\phi$-dependence of magnetoresistance of $Nd_{0.775}Sr_{0.225}NiO_2$. Fits based on Eq. 1 and 2 are shown in black dashed lines.

# Supplementary Materials for

## Rare-Earth Control of the Superconducting Upper Critical Field in Infinite-Layer Nickelates

Bai Yang Wang[1,2]*†, Tiffany C. Wang[2,3]*†, Yu-Te Hsu[4], Motoki Osada[2,5], Kyuho Lee[1,2], Chunjing Jia[2], Caitlin Duffy[4], Danfeng Li[2], Jennifer Fowlie[2,3], Malcolm R. Beasley[3], Thomas P. Devereaux[2,5], Ian R. Fisher[2,3], Nigel E. Hussey[4,6], Harold Y. Hwang[2,3]*

**Affiliations:**

[1]Stanford Institute for Materials and Energy Sciences, SLAC National Accelerator Laboratory; Menlo Park, CA 94025, United States.

[2]Department of Physics, Stanford University; Stanford, CA 94305, United States.

[3]Department of Applied Physics, Stanford University; Stanford, CA 94305, United States.

[4]High Field Magnet Laboratory (HFML-EMFL) and Institute for Molecules and Materials, Radboud University; Toernooiveld 7, 6525 ED Nijmegen, Netherlands.

[5]Department of Materials Science and Engineering, Stanford University; Stanford, CA 94305, United States.

[6]H. H. Wills Physics Laboratory, University of Bristol; Tyndall Avenue, Bristol BS8 1TL, United Kingdom.

*Corresponding author. Email: bwang87@stanford.edu; catwang@stanford.edu; hyhwang@stanford.edu.

†Both authors contributed equally.

**Materials and Methods**

Materials

Thin film nickelate samples (~7 nm thick) were grown on SrTiO$_3$ substrates using pulsed-laser deposition under growth and reduction conditions previously reported (*2, 3, 34*).

Methods

For angular magnetoresistance measurements, the samples were contacted using wire-bonded aluminum wires in a 6-point Hall bar geometry. The 50% resistive criterion is applied to both temperature-sweep and field-sweep data to determine $T_c$ and $H_{c2}$. For high field measurements, transport measurements were performed using the standard four-probe lock-in technique, with an alternating current excitation between 1 and 10 µA. Continuous magnetic fields up to 35 T were generated using a Bitter magnet, coupled with a He-3 cryostat, at the High Field Magnet Laboratory at Nijmegen, the Netherlands. The magnetic field angle was varied *in-situ* using a customized rotator stage.

**Supplementary Text**

High field magnetoresistance properties of additional La- and Pr- samples

We show the magnetoresistance for additional La$_{1-x}$Sr$_x$NiO$_2$, Pr$_{1-x}$Sr$_x$NiO$_2$, and Nd$_{1-x}$Sr$_x$NiO$_2$ (*12*) samples (fig. S1). Across Sr doping, both La- and Pr-nickelates require magnetic field up to 35 T (or beyond) to fully suppress superconductivity. In contrast, except for $x = 0.175$, a 20 T magnetic field is sufficient to bring the Nd-nickelates into the normal state. In addition, as Sr doping is increased, the field scales at which the superconductivity is destroyed converge between the two field orientations and eventually cross for $x = 0.225$.

Doping dependence of $H_{c2}$ for all three nickelate variants

Here we show the $H_{c2}$ data of Fig. 1F in a decompressed format to more clearly illustrate their doping dependences (fig. S2). For each variant, the bottom panel shows the corresponding superconducting dome traced out by the $T_{c0}$ values of the four dopings investigated. For each doping, the corresponding $H_{c2\perp}$ and $H_{c2\parallel}$ are plotted in a pair of panels, marked by the black arrows. For all three nickelate variants and across all measured dopings, near $T_{c0}$, we consistently observe the linear temperature dependence for $H_{c2\perp}$ (top panel of each pair) and $(1-T/T_{c0})^{1/2}$ temperature dependence for $H_{c2\parallel}$ (bottom panel of each pair).

Angular magnetoresistance properties of additional samples

The reported anomalous angular magnetoresistance is seen in all 16 measured Nd-nickelate samples and absent in all 7 measured La- and Pr-nickelates. We show the temperature and field dependence of the angular magnetoresistance of additional representative nickelate samples of all three $R$-variants in figs. S3 to S5. While there are sample to sample variations, the overall angular magnetoresistance behavior and doping evolution are qualitatively consistent. For example, while only two of the three $x = 0.225$ samples exhibit higher resistance at $\theta = 0$, the substantial impact of the EMP effect is evident in all three samples. The same fittings for the $\theta$- and $\phi$-dependence (Eqs. 1 and 2 of the main text) are shown as black dashed lines. In contrast, the La- and Pr-nickelates all follow the expected AOD dominant $\theta$-dependence and the vortex-driven $C_2$ $\phi$-dependence.

Vortex origin of anomalous angular magnetoresistance

Anomalous $\theta$- and $\phi$-dependent magnetoresistance features have been seen in cuprate superconductors, and attributed to the angular dependence of the vortex pinning mechanisms (23,24). For $\theta$-dependence, a symmetric pair of sharp resistance peaks is observed when the field orientation slightly deviates from the in-plane direction, which is a result of the introduction of mobile inter-layer segments of the vortices. In addition, a small resistance dip is observed for the $H_\perp$ orientation, which is due to vortices locking to vertical defects, such as domain or anti-phase boundaries (23). For $\phi$-dependence, a two-fold symmetric magnetoresistance pattern is seen, since the Lorentz force acting on the vortices has a two-fold $\phi$-dependence. Additionally, in YBa$_2$Cu$_3$O$_{7-x}$, a four-fold clover leaf pattern has been observed, which is attributed to the additional vortex pinning from orthorhombic domain boundaries (23,24).

However, our observed anomalous angular magnetoresistance cannot be explained by vortex pinning mechanisms. First, previous observations in the cuprates are sensitive to the measurement current density and are only visible in the presence of large current density (~8E+5 A/cm$^2$) and small magnetic field (6 T relative to ~70 K $T_{c0}$) (23). In contrast, our observations are seen in small current density (~40 A/cm$^2$) and large magnetic field (~$H_{c2}$). To examine this further, we studied the angular magnetoresistance dependence on measurement current. As shown in fig. S6, we vary the current density by up to 3 orders of magnitude and see no qualitative changes to either the $\theta$- or $\phi$-dependence. A further inconsistency with the previous cuprate observations is found in the angular position of the anomalous features in the $\theta$-dependence. While they only occur near the principle axes in the cuprates, the angular modulations reported in this work are predominantly between the principle axes. Considering these distinctions, we conclude that vortex dynamics are not the origin of our observations.

Crystal field calculations of rare-earth 4f moments

We investigate the crystal field splitting of the 4f levels through numerical calculations. We treat the static tetragonal crystal field as a perturbation Hamiltonian, described as:

$$H_{CEF} = B_2^0 O_2^0 + B_4^0 O_4^0 + B_6^0 O_6^0 + B_4^4 O_4^4 + B_6^4 O_6^4,$$

where $B_n^m$'s are the crystal field parameters and $O_n^m$'s are the Steven Operators. The five terms are the only non-zero terms given the tetragonal symmetry of the unit cell (35). While the decomposition of Steven operators O are well tabulated, the crystal field parameters $B_n^m$ are dependent on the exact distribution of the surrounding charge within the unit cell. For the $B_n^m$ calculation, we adopt the point charge approximation, as we are primarily interested in the energy hierarchy of the split levels (35).

To obtain the electron density distributions, we implement density functional theory (DFT) calculations with Perdew-Burke-Ernzerhof (26) exchange-correlation functional using the Vienna *Ab initio* Simulation Package (VASP) (27). The pseudopotentials for Nd and Pr are taken as the $Nd^{3+}$ and $Pr^{3+}$ with the 4f electrons in the core. We calculate the electron density distributions associated with R 5d and 6s orbitals, O 2p orbitals and Ni 3d and 4s orbitals, totaling 33 electrons per unit cell (11 electrons for R, and 10 electrons for Ni and 6 electrons for each O in the pseudopotentials). Using Bader charge analysis (28, 36) we dissect the total electron density distribution into separate regions and assign each region to the corresponding ions. We also treat the positive ionic charges as point charges localized at each ionic site. Adding the two, we obtain the effective valence of Ni, O, and Nd to be: $0.7^+$, $1.31^-$, and $1.93^+$, localized at each ionic site. Then $B_n^m$ can be calculated (*35*) according to:

$$B_n^m = \left(\sum_i \frac{4\pi}{2n+1} q_i \frac{Z_{nm}(\theta_i,\phi_i)}{R_i^{n+1}}\right) \eta_n \langle r^n \rangle,$$

where the summation is over all considered point charges, $R_i$ is the distance between the point charge and the Nd ion, and $\eta_n = \alpha_J/\beta_J/\gamma_J$ for $n$ = 2/4/6. The $Z_{nm}$, $\alpha_J/\beta_J/\gamma_J$, and $\langle r^n \rangle$ are previously tabulated (*35*). After diagonalizing the Hamiltonian, we find that the $Pr^{3+}$ $^3H_4$ 9-fold degeneracy and $Nd^{3+}$ $^4I_{9/2}$ 10-fold degeneracy are split into 7 and 5 Stark levels, respectively. The resulting levels assuming a 3.33 Å $c$-lattice constant are listed in Table S1, with their numbering in correspondence to Fig. 3 in the main text. In particular, the ground state of the split $Pr^{3+}$ 4$f$ levels is a singlet, as emphasized in the main text.

We can also calculate the easy axis anisotropy by examining $\langle g_J \mu_B J_{x/z} \rangle$ of the ground state levels in response to an external field along the $z$- or $x$-axis: $g_J \mu_B H_{x/z}$. Here $g_J$ is the Landé g-factor, $\mu_B$ is the Bohr magneton and $J_{x/z}$ is the total angular momentum operator projection along the $x$ or $z$ axis. For both Pr- and Nd-nickelates, we find a larger response when field is applied in-plane as shown in fig. S7, indicating an easy-plane anisotropy. This is expected as the ground state of the split 4$f$ levels has a dominant $|m = 0\rangle$ or $|m = \pm\frac{1}{2}\rangle$ component. The same crystal field splitting result can be obtained from symmetry arguments as well. For example, under a tetragonal symmetry, the $Pr^{3+}$ 4$f$ levels can be decomposed to 5 irreducible representations: $A_{1g}$, $A_{2g}$, $B_{1g}$, $B_{2g}$, $E_g$. They correspond to 5 nondegenerate energy levels and 2 doubly degenerate levels, consistent with previous reports (*37*) and our numerical calculation results.

Dependence of azimuthal angular magnetoresistance on measurement current direction

As discussed in the main text, the $C_2$ symmetric $\phi$-dependent magnetoresistance seen in all nickelate samples can be attributed to vortex dynamics induced by the measurement current, while the $C_4$ symmetric response is attributed to the Nd $4f$ electrons in the underlying tetragonal crystal symmetry. Therefore, we expect the $C_2$ symmetric response to rotate with the current direction and the $C_4$ symmetric response to remain locked onto the crystalline axes. We perform such a test on a $Nd_{0.8}Sr_{0.2}NiO_2$ and a $Pr_{0.84}Sr_{0.16}NiO_2$ sample by sourcing current along the [110] direction. As shown in fig. S9, the clover leaf pattern of the Nd-nickelate sample retains the same orientation as in Fig. 4, while the 'figure-of-eight'-shaped pattern of the Pr-nickelate sample is rotated by 45 degrees.

Indications of Nd antiferromagnetic order

We find indirect evidence for antiferromagnetic order of the $4f$ moments in the doping evolution of the angular magnetoresistance measurements. Shown in fig. S10 are representative polar angle dependence data for $x = 0.15$, 0.175, 0.2 (sample #1), and 0.225 (sample #1) Sr doping at 2 K. A gradual but clear transition from an AOD dominant '∞'-shaped pattern to an EMP dominant 'figure-of-eight'-shaped pattern is seen as doping is increased. Such an enhancement of the magnetic response of the $4f$ moments might seem counterintuitive, given the substitution of non-magnetic $Sr^{2+}$ into the $Nd^{3+}$ lattice. However, in the case of antiferromagnetism, non-magnetic defects can amplify the suppression of superconductivity by magnetic ordering (*29*), with the defects serving as effective magnetic scatterers due to the local imbalance between the up- and down-spin sublattice. In this sense, the observed doping dependence is consistent with antiferromagnetic ordering of the Nd 4*f* moments. Ultimately, a direct probe of Nd magnetism, although challenging for these thin films at very low temperatures, would be extremely insightful.

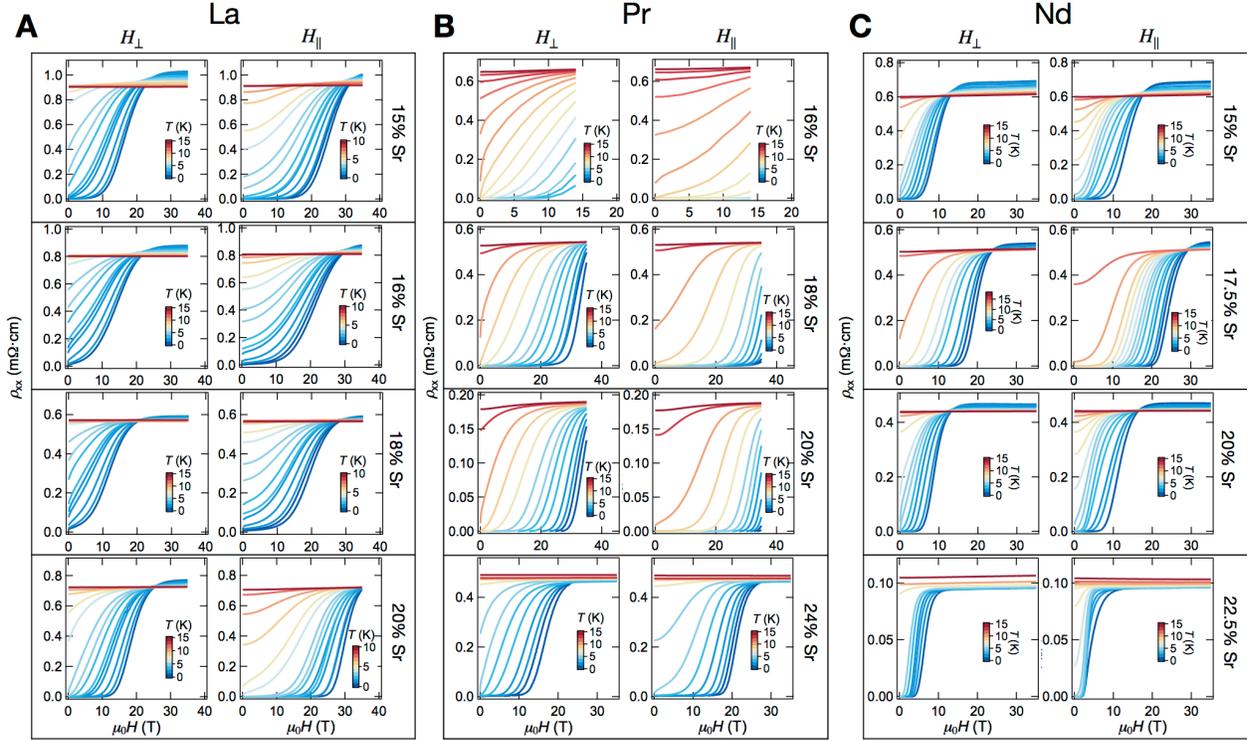

**Fig. S1.**

**Doping dependence of the high field magnetoresistance in $R_{1-x}Sr_xNiO_2$.** (**A - C**) Doping dependence of the $La_{1-x}Sr_xNiO_2$, $Pr_{1-x}Sr_xNiO_2$, and $Nd_{1-x}Sr_xNiO_2$ (*12*) magnetoresistance at temperatures ranging from 0.34 K to 18 K. For La-nickelates, $x$ = 0.15, 0.16, 0.18, and 0.2; for Pr-nickelates, $x$ = 0.16, 0.18, 0.2, and 0.24; and for Nd-nickelates, $x$ = 0.15, 0.175, 0.2, and 0.225. The two columns of panels correspond to $H_\perp$ (left) and $H_\parallel$ (right) field orientations.

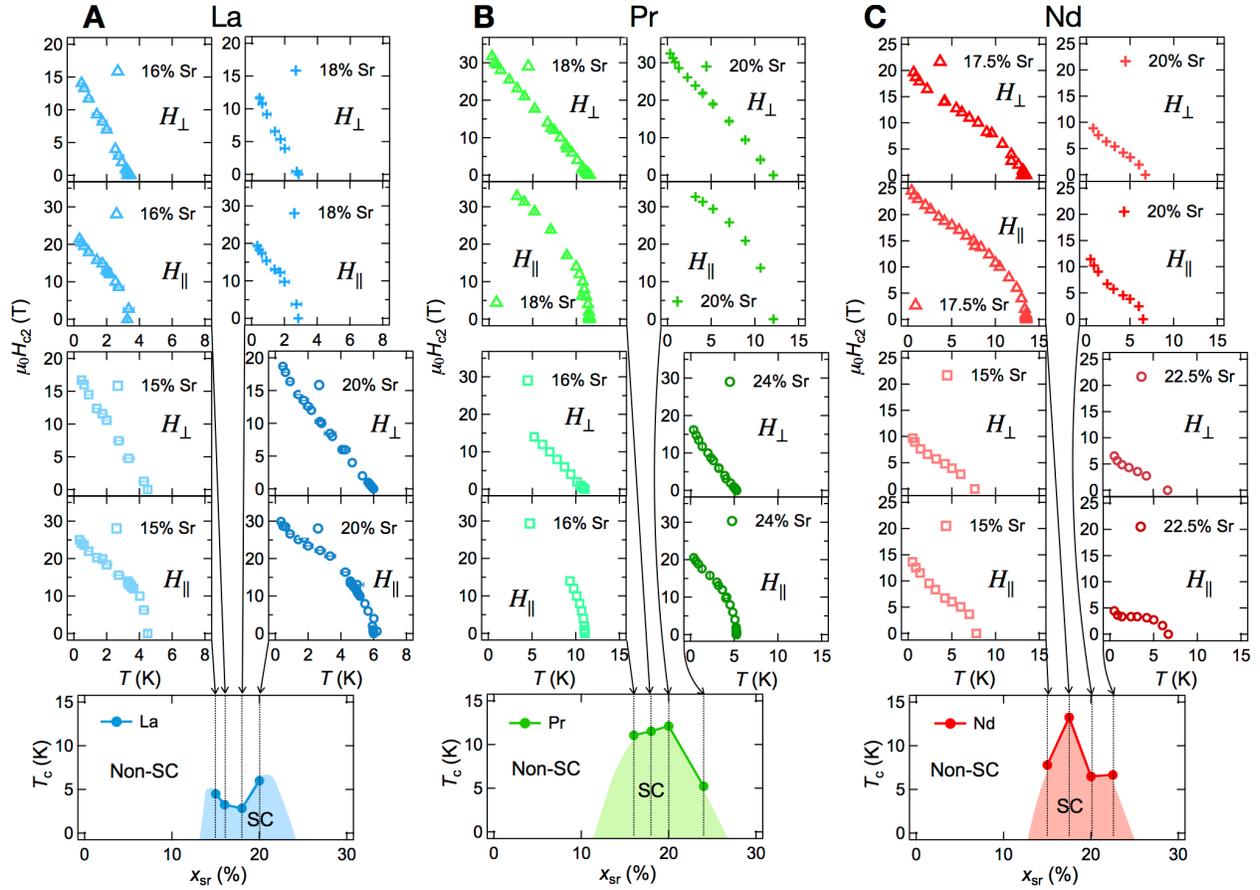

**Fig. S2.**

**Doping dependence of $H_{c2}$ in $R_{1-x}Sr_xNiO_2$.** (A - C) $H_{c2}(T)$ data shown in fig. 1F of $La_{1-x}Sr_xNiO_2$, $Pr_{1-x}Sr_xNiO_2$, and $Nd_{1-x}Sr_xNiO_2$ (*12*). The doping of each pair is marked, by a black arrow, onto the corresponding superconducting domes shown in the three bottom panels.

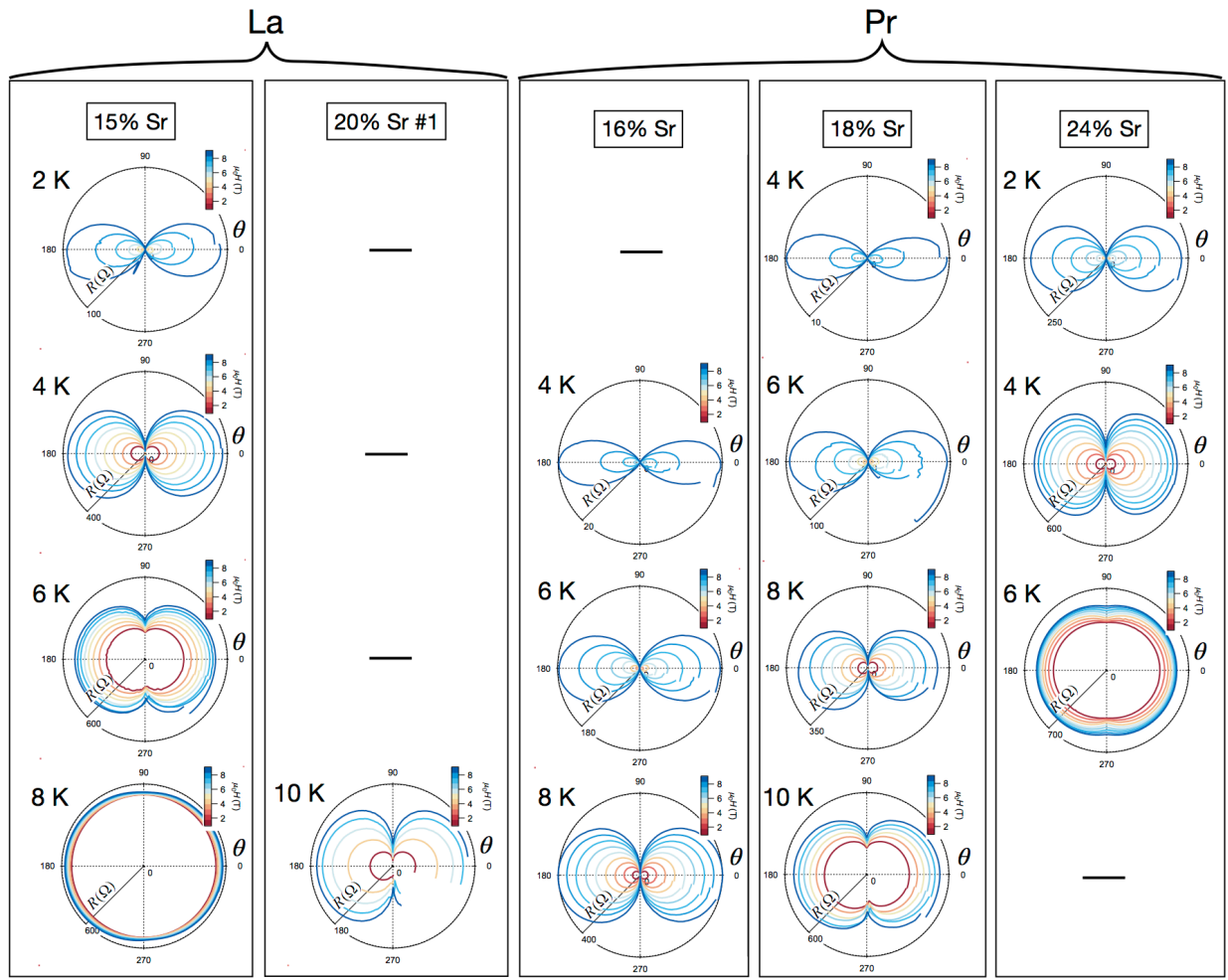

**Fig. S3**

**Polar angular magnetoresistance of La- and Pr-nickelates.** Temperature and magnetic field dependence of the polar angular magnetoresistance of $La_{1-x}Sr_xNiO_2$ for $x = 0.15$ and $0.2$ and $Pr_{1-x}Sr_xNiO_2$ samples for $x = 0.16$, $0.18$, and $0.24$. Each column plots the data set of a separate sample. The different curves within each panel correspond to external field of 1 to 9 T (in 1 T increments), with larger resistance at larger field. The measurement temperature corresponding to each panel is shown in the top left corner.

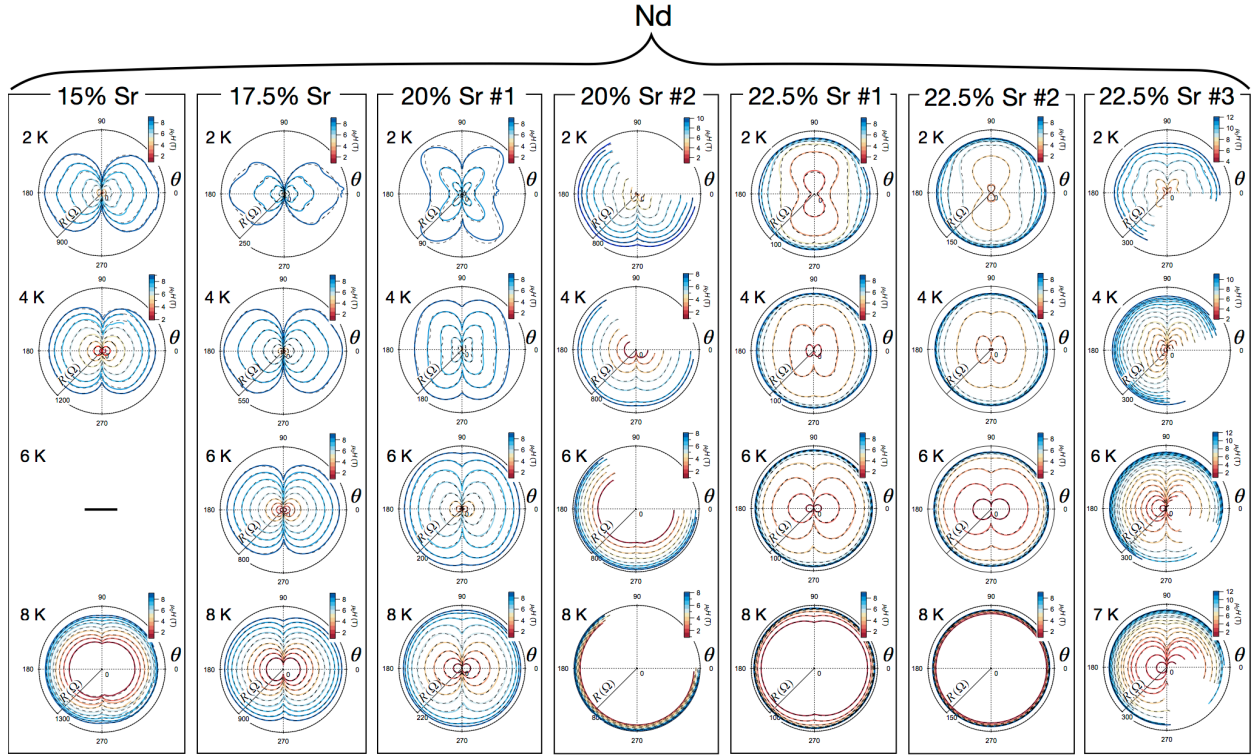

**Fig. S4.**

**Doping dependence of the polar angular magnetoresistance in $Nd_{1-x}Sr_xNiO_2$.** Doping dependence of the polar angular magnetoresistance at 2, 4, 6, 7 and 8 K for $x$ = 0.15, 0.175, 0.2, and 0.225. Each column plots the data set of a separate sample. For $x$ = 0.225, sample #1 corresponds to that in the main text Fig. 2 and 4. The different curves within each panel correspond to external field of 1 to 9 T, with larger resistance at larger field. The temperature corresponding to each panel is shown at its top left corner. Fits based on Eq. 1 in the main text are shown in black dashed lines.

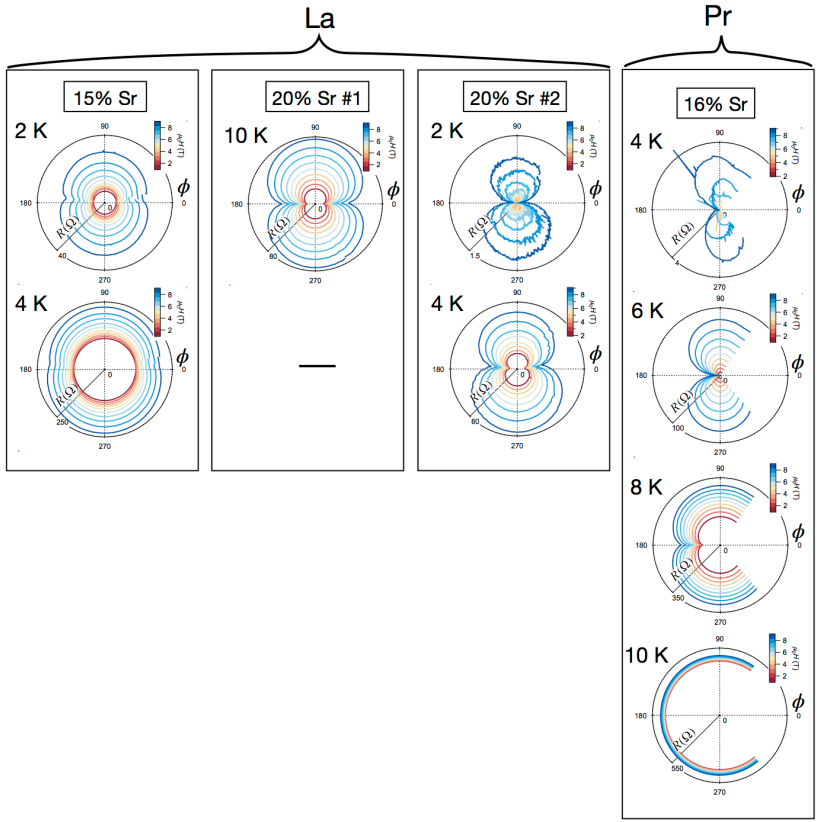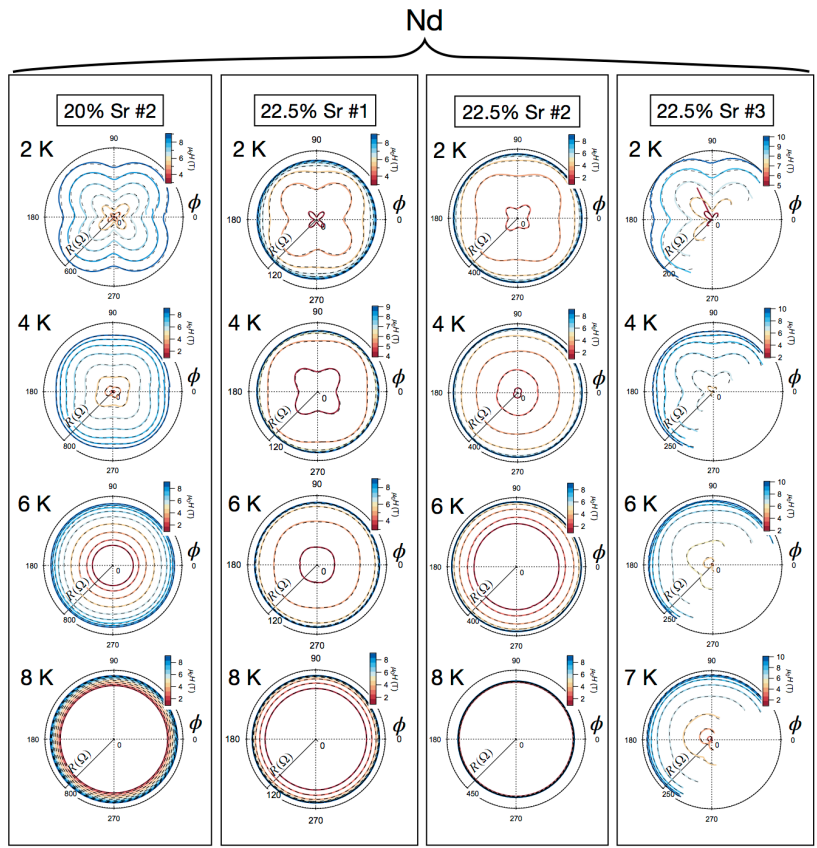

**Fig. S5.**

**Azimuthal angular magnetoresistance of nickelates across A-site variants.** Temperature and magnetic field dependence of polar angular magnetoresistance. Each column represents a separate sample. For $Nd_{0.775}Sr_{0.225}NiO_2$, the three samples correspond to those in fig. S4. Different curves within each panel correspond to external field of 1 to 9 T (in 1 T increments), with larger resistance at larger field. The temperature corresponding to each panel is shown at its top left corner. Fits to the Nd-nickelate data based on Eq. 2 in the main text are shown in black dashed lines.

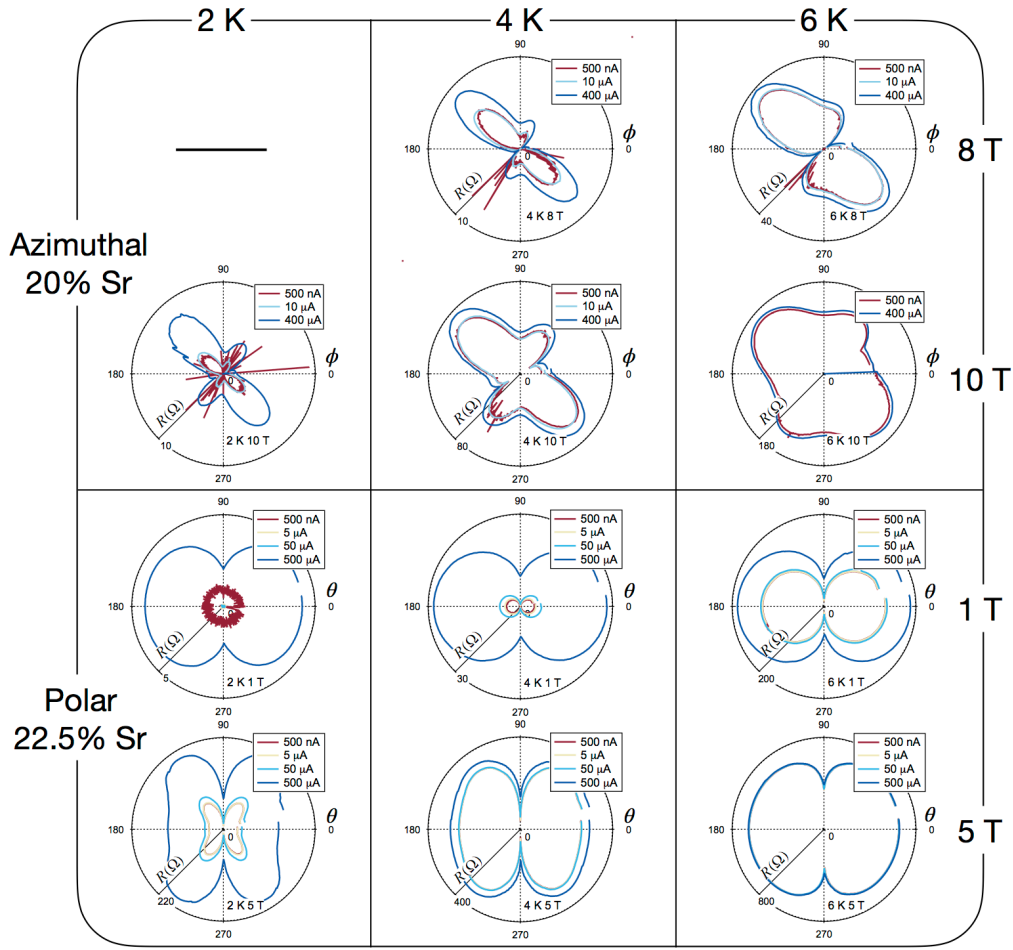

**Fig. S6.**

**Current dependence of the anomalous angular magnetoresistance.** Current dependence of the $\phi$-dependent magnetoresistance of a $Nd_{0.8}Sr_{0.2}NiO_2$ sample at temperatures 2, 4, and 6 K and fields 8 and 10 T (upper panels), and the $\theta$-dependent magnetoresistance of a $Nd_{0.775}Sr_{0.225}NiO_2$ sample at temperatures 2, 4, and 6 K and fields 1 and 5 T (lower panels). Different curves within each panel are associated with measurement currents ranging from 0.5 μA to 500 μA, corresponding to ~2 A/cm$^2$ to 2 kA/cm$^2$.

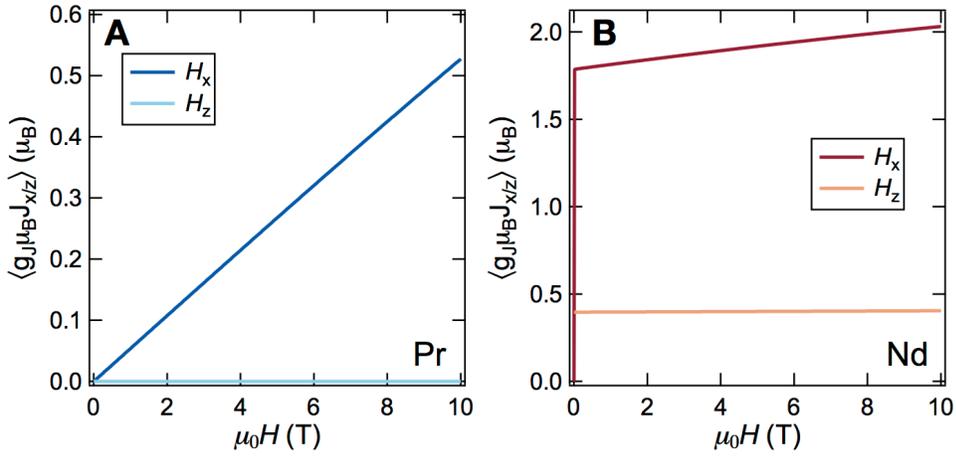

**Fig. S7.**

**Calculated magnetic easy axis.** (**A** and **B**) Effective magnetic moment of $Pr^{3+}$ and $Nd^{3+}$ in the presence of external in-plane and out-of-plane magnetic field for the ground state configuration.

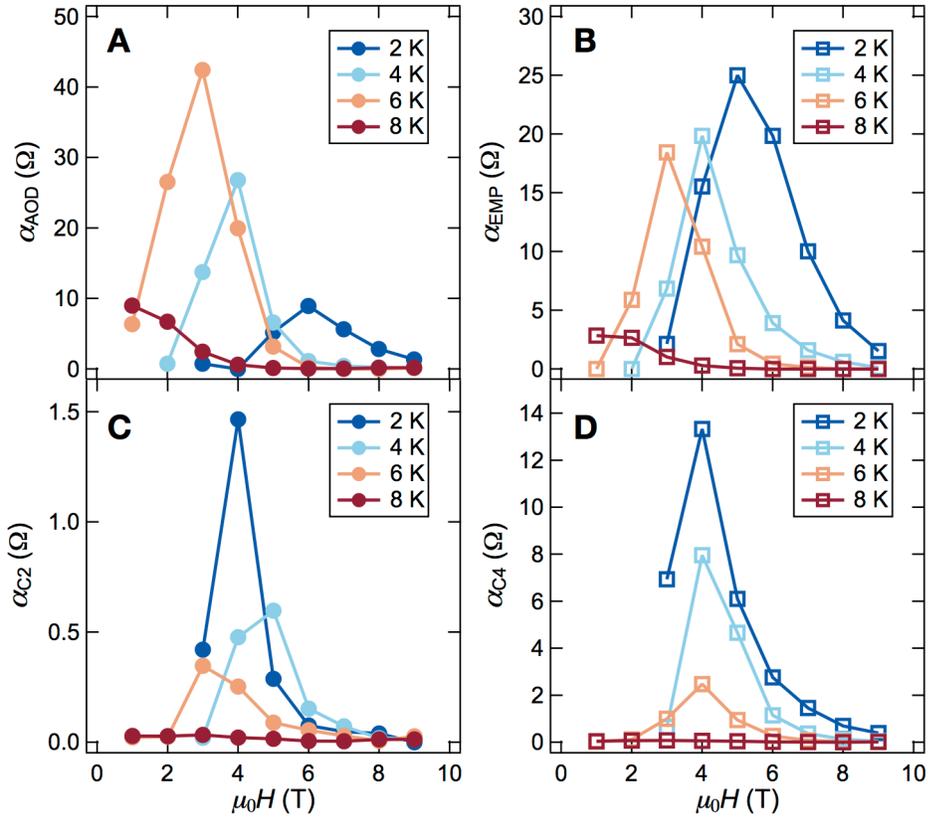

**Fig. S8.**

**Temperature and field dependence of the fitting parameters.** (**A** and **B**) Temperature and field dependence of the fit coefficients corresponding to the AOD and EMP terms in the polar angular dependence, respectively. (**C** and **D**) Temperature and field dependence of the fit coefficients corresponding to the $C_2$ and $C_4$ symmetric terms of the azimuthal angular dependence, respectively.

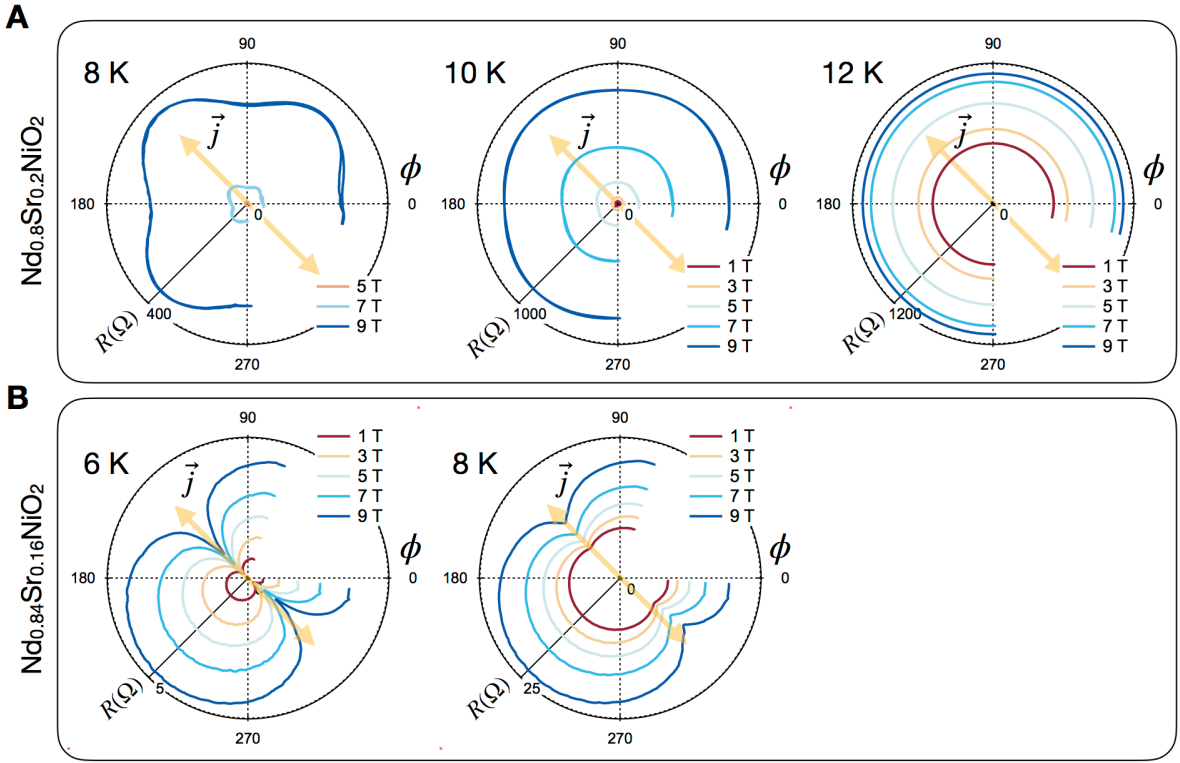

**Fig. S9.**

**Current direction dependence of the $\phi$-dependent magnetoresistance.** (**A**) $\phi$-dependent magnetoresistance of a $Nd_{0.8}Sr_{0.2}NiO_2$ sample at temperatures 8, 10, and 12 K and fields from 1 to 9 T. (**B**) $\phi$-dependent magnetoresistance of a $Pr_{0.84}Sr_{0.16}NiO_2$ sample at temperatures 6 and 8 K and fields from 1 to 9 T. The direction of the measurement current is illustrated by the yellow arrow in both panels. For this specific measurement, $\phi = 0$ corresponds to the direction of the crystal $a$-axis.

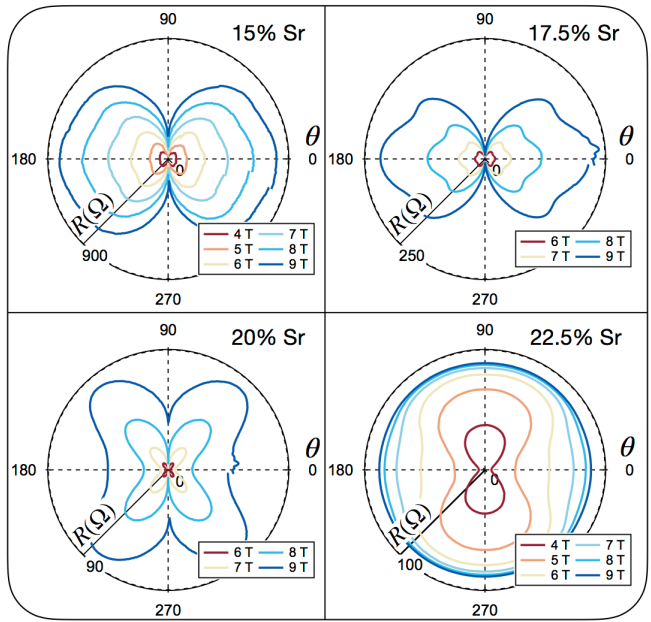

**Fig. S10.**

**Indications of Nd 4*f* moment antiferromagnetic order in the doping dependence.** Doping evolution of the $(Nd,Sr)NiO_2$ $\theta$-dependent magnetoresistance at 2 K, with the doping level given at the top right corner of each panel.

| Pr Levels | Pr $\|m\rangle$ Components | Nd Levels | Nd $\|m\rangle$ Components |
|---|---|---|---|
| $L_0$ | $\|0\rangle$ | $L_0$ | $0.15\left\|\frac{9}{2}\right\rangle - 0.98\left\|\frac{1}{2}\right\rangle + 0.12\left\|-\frac{7}{2}\right\rangle$ |
| $L_1$ | $0.99\|1\rangle - 0.1\|-3\rangle$ | $L_1$ | $-0.15\left\|-\frac{9}{2}\right\rangle + 0.98\left\|-\frac{1}{2}\right\rangle - 0.12\left\|\frac{7}{2}\right\rangle$ |
| $L_2$ | $-0.99\|-1\rangle + 0.1\|3\rangle$ | $L_2$ | $-0.09\left\|\frac{5}{2}\right\rangle + 0.99\left\|-\frac{3}{2}\right\rangle$ |
| $L_3$ | $0.71\|2\rangle - 0.71\|-2\rangle$ | $L_3$ | $0.09\left\|-\frac{5}{2}\right\rangle - 0.99\left\|\frac{3}{2}\right\rangle$ |
| $L_4$ | $0.71\|2\rangle + 0.71\|-2\rangle$ | $L_4$ | $-0.09\left\|\frac{3}{2}\right\rangle - 0.99\left\|-\frac{5}{2}\right\rangle$ |
| $L_5$ | $0.73\|3\rangle - 0.07\|1\rangle + 0.07\|-1\rangle - 0.68\|-3\rangle$ | $L_5$ | $-0.99\left\|\frac{5}{2}\right\rangle - 0.09\left\|-\frac{3}{2}\right\rangle$ |
| $L_6$ | $-0.68\|3\rangle - 0.07\|1\rangle - 0.07\|-1\rangle - 0.73\|-3\rangle$ | $L_6$ | $-0.12\left\|\frac{9}{2}\right\rangle + 0.1\left\|\frac{1}{2}\right\rangle + 0.99\left\|-\frac{7}{2}\right\rangle$ |
| $L_7$ | $0.71\|4\rangle - 0.71\|-4\rangle$ | $L_7$ | $-0.12\left\|-\frac{9}{2}\right\rangle + 0.1\left\|-\frac{1}{2}\right\rangle + 0.99\left\|\frac{7}{2}\right\rangle$ |
| $L_8$ | $0.71\|4\rangle + 0.71\|-4\rangle$ | $L_8$ | $0.98\left\|-\frac{9}{2}\right\rangle + 0.16\left\|-\frac{1}{2}\right\rangle + 0.10\left\|\frac{7}{2}\right\rangle$ |
| | | $L_9$ | $0.98\left\|\frac{9}{2}\right\rangle + 0.16\left\|\frac{1}{2}\right\rangle + 0.10\left\|-\frac{7}{2}\right\rangle$ |

**Table S1.**

**CEF split 4f levels.** Compositions of the 4f eigenstates under the tetragonal crystal field assuming a 3.33 Å c-lattice constant. Here $|m\rangle$ refers to the originally degenerate 4f atomic states, with $m$ denoting the projected angular momentum quantum number. The $L_x$ numberings correspond to Figs. 3, B and C in the main text.